\documentclass[prl,aps,twocolumn,showpacs]{revtex4}
\usepackage{epsfig}
\usepackage{bm}

\begin{document}

\title{Nonlinear and chaotic resonances in solar activity}

\author{\small  A. Bershadskii}
\affiliation{\small {ICAR, P.O.B. 31155, Jerusalem 91000, Israel}}

\begin{abstract}
It is shown that, the wavelet regression detrended  fluctuations of the 
monthly sunspot number for 1749-2009 years exhibit strong periodicity with 
a period approximately equal to 3.7 years. The wavelet regression method 
detrends the data from the approximately 11-years period. Therefore, it is 
suggested that the one-third subharmonic resonance can be considered as a 
background for the 11-years solar cycle. It is also shown that the broad-band 
part of the wavelet regression detrended fluctuations spectrum exhibits an 
exponential decay that, together with the positive largest Lyapunov exponent, are 
the hallmarks of chaos. Using a complex-time analytic approach the rate of the 
exponential decay of the broad-band part of the spectrum has been theoretically 
related to the Carrington solar rotation period.  Relation of the driving period 
of the subharmonic resonance (3.7-years) to the active longitude flip-flop 
phenomenon, in which the dominant part of the sunspot activity changes the 
longitude every 3.7 years on average, has been briefly discussed.

\end{abstract}


\maketitle

\begin{figure} \vspace{-0.5cm}\centering
\epsfig{width=.45\textwidth,file=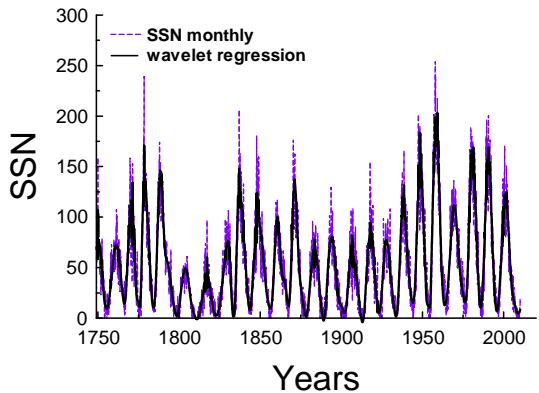} \vspace{-5cm}
\caption{The monthly sunspot number (dashed line) for the period 1749-2009 years. 
The data were taken from Ref. \cite{belg}. The solid curve 
(trend) corresponds to a wavelet (symmlet) regression of the data. }
\end{figure}

The solar activity is chaotic but has a well-defined mean period of
about 11 years. The 11-year cycle is well known for more than a century and a half. 
Despite this, nature of the 11-year cycle is still unknown. Most of the regression 
methods are linear in responses and statistical analyses of the experimental sunspot 
data was dominated by linear stochastic methods, while it was recently rigorously 
shown in Ref. \cite{pn} that a nonlinear dynamical mechanism (presumably a driven nonlinear 
oscillator) determines the sunspot cycle. Figure 1 shows 
the monthly sunspot number (dashed line) for the period 1749-2009 years 
(the data were taken from Ref. \cite{belg}). The solid curve 
(trend) corresponds to a wavelet (symmlet) regression of the data (cf. Ref. \cite{o}). 
Figure 2 shows corresponding detrended fluctuations, which produce 
a statistically stationary set of data. At the nonlinear nonparametric wavelet regression 
one chooses a relatively small number of wavelet 
coefficients to represent the underlying regression function. A threshold method is used to keep or 
kill the wavelet coefficients. In this case, in particular, the Universal (VisuShrink) thresholding 
rule with a soft thresholding function was used. At the wavelet 
regression the demands to smoothness of the function being estimated are relaxed considerably in comparison 
to the traditional methods. 
Figure 3 shows a spectrum of the wavelet regression detrended data 
calculated using the maximum entropy method (because it provides an optimal spectral
resolution even for small data sets). In Fig. 3 one can see a well defined peak corresponding 
to period $\sim$ 3.7 years. The wavelet regression method 
detrends the data from the approximately 11-years period (cf. Fig. 1). Therefore, it is plausible 
that the one-third subharmonic resonance \cite{nm} can be considered as a background for 
the 11-years solar cycle: $11/3.7 \simeq 3$. Indeed, it is known \cite{nocera} that 
interaction of the Alfven waves (generated in a highly magnetized plasma by a cavity's moving boundaries) 
with slow magnetosonic waves can be described using Duffing oscillators (see also Refs. \cite{pn},\cite{pl}). 
Let us imagine a forced 
excitable system with a large amount of loosely coupled degrees of freedom schematically 
represented by Duffing oscillators (which has become a classic model for analysis of 
nonlinear phenomena and can exhibit both deterministic and chaotic behavior \cite{nm},\cite{ph} 
depending on the parameters range) with a wide range of the natural frequencies $\omega_0$ :

$$
\ddot{x} + \omega_0^2 x +\gamma \dot{x} +\beta x^3 = F \sin\omega t    \eqno{(1)}
$$
where $\dot{x}$ denotes the temporal derivative of $x$, $\beta$ is the strength of nonlinearity, and 
$F$ and $\omega$ are characteristic of a driving force. It is known (see for instance Ref. \cite{nm}) 
that when $\omega \approx 3\omega_0$ and $\beta \ll 1$ the equation (1) has a resonant solution 
$$
x(t) \approx a \cos\left(\frac{\omega}{3}t + \varphi \right) + \frac{F}{(\omega^2-\omega_0^2)} 
\cos \omega t   \eqno{(2)}
$$
where the amplitude $a$ and the phase $\varphi$ are certain constants. 
This is so-called one-third subharmonic resonance with the driving frequency 
$\omega$ corresponding approximately to 3.7 years period (the peak in Fig. 3 
corresponds to the second term in the right-hand side of the Eq. (2) while the first term 
has been detrended). 
For the considered system of the oscillators an effect of synchronization can take place
and, as a consequence of this synchronization, the characteristic peaks in the spectra 
of partial oscillations coincide \cite{nl}. 
It can be useful to note, for the solar activity modeling, that the odd-term subharmonic resonance 
is a consequence of the reflection symmetry of the natural nonlinear oscillators 
(invariance to the transformation $x \rightarrow -x$). \\

\begin{figure} \vspace{-1cm}\centering
\epsfig{width=.45\textwidth,file=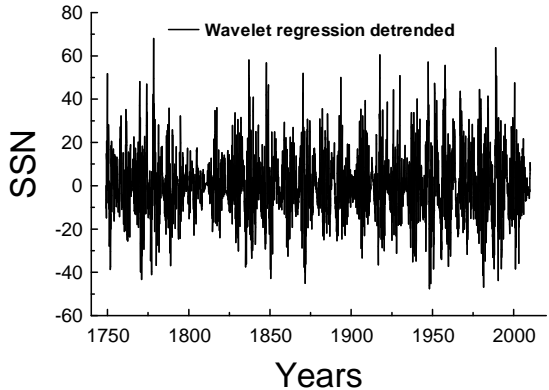} \vspace{-5cm}
\caption{The wavelet regression detrended fluctuations from the data shown in Fig. 1.}
\end{figure}
\begin{figure} \vspace{-0.5cm}\centering
\epsfig{width=.45\textwidth,file=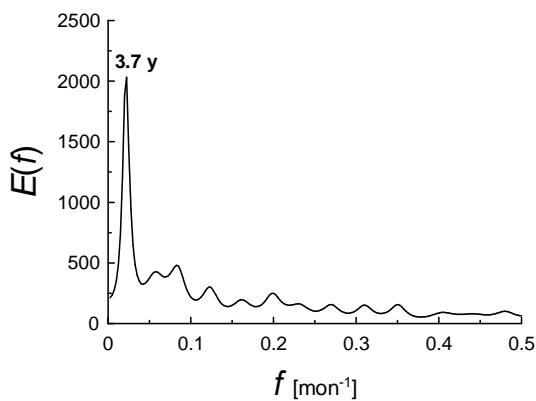} \vspace{-5cm}
\caption{Spectrum of the wavelet regression detrended fluctuations shown 
in Fig. 2.}
\end{figure}

\begin{figure} \vspace{-1cm}\centering
\epsfig{width=.45\textwidth,file=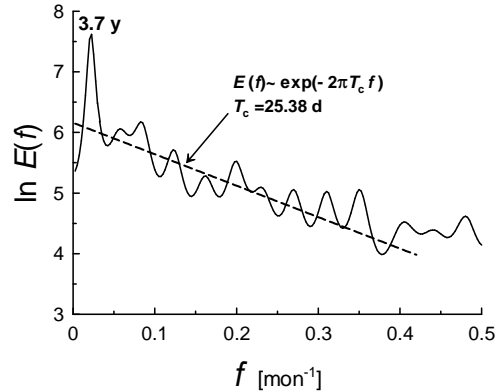} \vspace{-5cm}
\caption{The same as in Fig. 3 but in semi-logarithmical scales. The dashed straight line 
indicates an exponential decay.}
\end{figure}

In order to understand appearance of the $3.7$-years period let us represent the spectrum shown in Fig. 3 
in semi-logarithmical scales: figure 4. In these scales an exponential behavior corresponds to a straight 
line. It is known, that both stochastic and deterministic processes can result in the
broad-band part of the spectrum, but the decay in the
spectral power is different for the two cases. The exponential
decay indicates that the broad-band spectrum
for these data arises from a deterministic rather than a
stochastic process. For a wide class of deterministic
systems a broad-band spectrum with exponential
decay is a generic feature of their chaotic solutions 
Refs. \cite{oht}-\cite{fm}. A wavy exponential decay (see Fig. 4) is a characteristic of 
a chaotic behavior generated by {\it time-delay} differential equations \cite{fa}. 
A classic example of time-delay differential equation with chaotic solutions is the 
Mackey-Glass equation:
$$
\frac{du(t)}{dt} = \frac{0.2 \cdot u(t-\tau)}{(1+u(t-\tau)^{10})} - 0.1 \cdot u(t)  \eqno{(3)} 
$$
Figure 5 shows spectrum of a solution of this equation for the time-delay $\tau=30$. 
The dashed straight line indicates an exponential decay (cf. Fig. 4). 

In the dynamo models that have physically distinct source layers the 
finite time is required in order to transport magnetic flux from one layer 
to another \cite{ws}, it is especially significant for those dynamo models that have 
spatially segregated source regions for the poloidal and toroidal magnetic field components 
(such as, for instance, the Babcock-
Leighton dynamo mechanism \cite{chau}). In the global 
dynamo models that include meridional circulation the time delay related to the 
circulation should be comparable to global rotation period (see below).

Nature of the exponential decay of the power spectra
of the chaotic systems is still an unsolved mathematical
problem. A progress in solution of this problem
has been achieved by the use of the analytical continuation
of the equations in the complex domain (see, for 
instance, \cite{fm}). In this approach the exponential decay
of chaotic spectrum is related to a singularity in the
plane of complex time, which lies nearest to the real axis (see 
the insert in Fig. 5).
Distance between this singularity and the real axis determines
the rate of the exponential decay. For many interesting cases 
chaotic solutions are analytic in a finite strip around the real time axis. 
This takes place, for instance for attractors bounded in the real 
domain (the Lorentz attractor, for instance). 
In this case the radius of convergence of the Taylor series 
is also bounded (uniformly) at any real time.

Let us consider, for 
simplicity, solution $u(t)$ with simple poles only, and to define the Fourier 
transform as follows
$$
\tilde{u}(f) =(2\pi)^{-1/2} \int_{-T_e/2}^{T_e/2} dt~e^{-i 2\pi f t} u(t)  \eqno{(2)}
$$  
Then using the theorem of residues
$$
\tilde{u}(f) =i (2\pi)^{1/2} \sum_j R_j \exp (i 2\pi f x_j -|2\pi f y_j|)  \eqno{(3)}
$$
where $R_j$ are the poles residue and $x_j + iy_j$ are their location in the relevant half
plane, one obtains asymptotic behavior of the spectrum $E(f)= |\tilde{u}(f)|^2$ at large $f$
$$
E(f) \sim \exp (-4\pi |y_{min}|~ f)  \eqno{(4)}
$$
where $y_{min}$ is the imaginary part of the location of
the pole which lies nearest to the real axis. In the case of symmetric analytic strip with a width 
$\Delta= 2 |y_{min}|$:
$$
E(f) \sim \exp (-2\pi \Delta~f )  \eqno{(5)}
$$
(cf. the insert in Fig. 5). 

\begin{figure} \vspace{-1.2cm}\centering
\epsfig{width=.48\textwidth,file=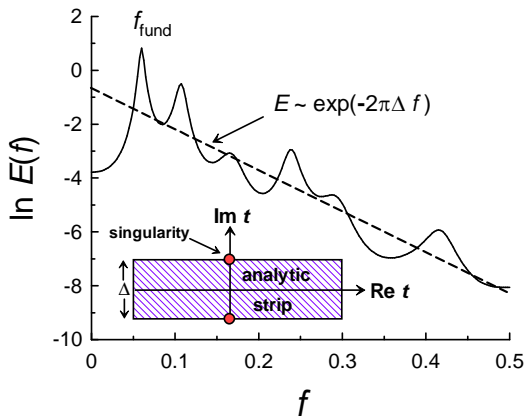} \vspace{-5cm}
\caption{Spectrum of the Mackey-Glass chaotic time-series. The dashed straight line 
indicates an exponential decay. The insert shows 
a sketch of corresponding complex-time plane.}
\end{figure}
\begin{figure} \vspace{-0.5cm}\centering
\epsfig{width=.45\textwidth,file=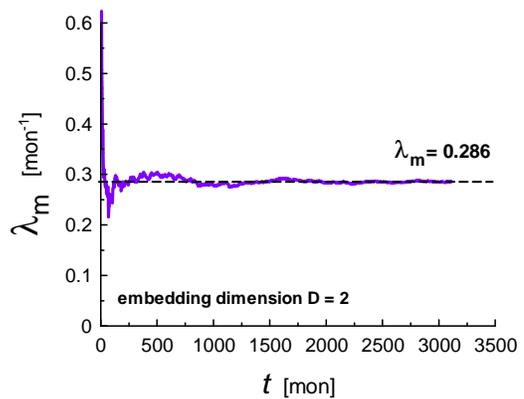} \vspace{-4.5cm}
\caption{The pertaining average maximal Lyapunov exponent 
at the pertaining time, calculated for the same data 
as those used for calculation of the spectrum (Figs. 3 and 4). 
The dashed straight line indicates convergence to a positive value.}
\end{figure}
The chaotic spectrum provides two different characteristic
time-scales for the chaotic system: a period corresponding to
fundamental frequency of the system, $f_{fund}$, and a period
corresponding to the exponential decay rate, $2\pi \Delta$ 
(cf. Eq. (5)). The fundamental period can be estimated
using position of the low-frequency peak (cf. Figs. 4 and 5), while the
exponential decay rate period $2\pi \Delta$ can be estimated
using the slope of the straight line of the broad-band part
of the spectrum in the semi-logarithmic representation. In the case 
of the global solar dynamo the width of the analytic strip $\Delta$ 
can be theoretically estimated using the Carrington solar rotation period: 
$\Delta \simeq T_c \simeq 25.38$ days. 
This period roughly corresponds to the solar rotation at a latitude of 26 deg, 
which is consistent with the typical latitude of sunspots (cf. Fig. 4).  \\

Additionally to the exponential spectrum (Fig. 4), let us check the chaotic 
character of the wavelet regression detrended 
fluctuations calculating the largest Lyapunov exponent: $\lambda_{max}$. A strong indicator for the presence 
of chaos in the examined time series is condition $\lambda_{max} >0$. If this is the case, then 
we have so-called exponential instability. Namely, 
two arbitrary close trajectories of the system will diverge apart exponentially, that is 
the hallmark of chaos. To calculate $\lambda_{max}$ we used a direct algorithm developed by 
Wolf et al. \cite{w}. Figure 6 shows 
the pertaining average maximal Lyapunov exponent at the pertaining time, calculated for the data set shown in 
Fig. 2. The largest Lyapunov exponent converges very 
well to a positive value $\lambda_{max} \simeq 0.286~ mon^{-1} > 0$.\\

It should be noted that the same period $\sim 3.7$ years was recently found for the so-called 
flip-flop phenomenon of the active longitudes in solar activity \cite{bu},\cite{bu2}. Sunspots 
are tend to pop up preferably in certain latitudinal domains and move toward the equator due to 
the 11-year cycle. Recently, strong indications of non-uniform {\it longitudinal} distribution of 
sunspots (active longitudes) was reported and analyzed in a dynamic frame related
to the mean latitude of sunspot formation, in which the active 
longitudes persist for the last eleven solar 11-years cycles 
(see Refs. \cite{bu},\cite{bu2} and references therein). At any given time, one of the two active 
longitudes (approximately $180^0$ apart) exhibits a stronger activity - dominance. Observed alternation 
of the active longitudes dominance in 3.7 years on average was called as flip-flop phenomenon 
\cite{bu}. It seems rather plausible that the observed flip-flop period and the period of the 
wavelet regression detrended fluctuations of solar activity (Fig. 4) have the same origin. In this vein, 
the observation \cite{bu},\cite{mh} that the period of the flip-flop phenomenon follows to 
variations of the real length of the sunspot cycle (which has the 11-years period on average only) 
supports the idea of the one-third subharmonic resonance as a background of the 11-years cycle of solar 
activity. \\

The author is grateful to SIDC-team, World Data Center for the Sunspot Index,
Royal Observatory of Belgium for sharing their data. 
A software provided by K. Yoshioka was used at the computations.

\end{document}